\documentstyle[11pt,epsf,fleqn]{elsart}

\newcommand{\be}[1]{\begin{equation} \label{(#1)}}
\newcommand{\ee}{\end{equation}}
\newcommand{\ba}[1]{\begin{eqnarray} \label{(#1)}}
\newcommand{\ea}{\end{eqnarray}}
\newcommand{\nn}{\nonumber}
\newcommand{\rf}[1]{(\ref{(#1)})}

\begin{document}
\begin{frontmatter}
\title{{\bf Towards a superformula for neutrinoless double beta decay}}
\author{H. P\"as\thanksref{e1}}, 
\author{M. Hirsch\thanksref{e2}},
\author{H.V. Klapdor--Kleingrothaus\thanksref{e3}},
\author{S. G. Kovalenko\thanksref{e4}}
\address{
Max--Planck--Institut f\"ur Kernphysik,\\ 
P.O.Box 10 39 80, D--69029 Heidelberg, Germany}
\thanks[e1]{E-mail: Heinrich.Paes@mpi-hd.mpg.de}
\thanks[e2]{Present Address: {\it Inst. de Fisica Corpuscular - C.S.I.C.
- Dept. de Fisica Teorica, Univ. de Valencia, 46100 Burjassot, Valencia, 
Spain}, \\
E-mail: mahirsch@flamenco.ific.uv.es}
\thanks[e3]{E-mail: klapdor@mickey.mpi-hd.mpg.de}
\thanks[e4]{On leave from {\it Joint Institute for 
Nuclear Research, Dubna, Russia},\\
E-mail: kovalen@nusun.jinr.dubna.su}
\begin{abstract}
A general Lorentz--invariant parameterization for the long-range 
part of the $0\nu\beta\beta$ decay rate is derived. Combined with 
the short range part this general parameterization in terms of 
effective $B-L$ violating couplings will allow it to extract the 
$0\nu\beta\beta$ limits on arbitrary lepton number violating theories.
Several new nuclear matrix elements appear in the general formalism 
compared to the standard neutrino mass mechanism. Some of these new 
matrix elements have never been considered before and are calculated 
within pn-QRPA. Using these, limits on lepton number violating 
parameters are derived from experimental data on $^{76}$Ge.
% insert abstract here
 \end{abstract}
\end{frontmatter}
\section{Introduction}
Double beta decay has been proven to be one of the most powerful 
tools to constrain $B-L$ violating physics beyond the standard model
\cite{Kla97}. In recent years, besides the most restrictive limit on 
the effective neutrino Majorana mass \cite{Kla97,HM97}, 
stringent 
constraints on several theories beyond the Standard Model such as
$R$--parity violating \cite{Hir95,Hir96,Paes98} as well as conserving 
\cite{Hir97} SUSY, leptoquarks \cite{Hir96b}, 
left--right symmetric models \cite{Hir96c} and compositeness 
\cite{Pan97,Tak97} have been derived (for a review see \cite{Kla97}).

While the neutrino mass limit is based on the well-known mechanism
exchanging a massive Majorana neutrino between two standard model $V-A$
vertices, the effective vertices appearing in the new contributions 
involve non--standard 
currents 
such as scalar, pseudoscalar and tensor 
currents. 

Thus we felt motivated to consider the neutrinoless double beta decay 
rate in a general framework, parameterizing the new physics contributions 
in terms of all effective low-energy currents allowed by Lorentz-invariance. 
Such an ansatz allows one to separate the nuclear physics part of double 
beta decay from the underlying particle physics model, and derive limits 
on arbitrary lepton number violating theories. 
The first step of this work, treating the long--range part, is presented 
here. Although the general decay rate is independent of the underlying 
nuclear physics model, to extract quantitative limits values for nuclear 
matrix elements are needed. First limits are derived using matrix elements 
calculated in proton-neutron (pn) QRPA, partially already available in 
the literature and partially calculated here for the first time.

\section{General Formalism}
We consider the long--range part of neutrinoless double beta decay with 
two vertices, which are pointlike at the Fermi scale, and exchange of a
light neutrino in between. The general Lagrangian can be written in terms of
effective couplings $\epsilon^{\alpha}_{\beta}$, which correspond to the
pointlike vertices at the Fermi scale so that Fierz rearrangement is 
applicable:
\be{1}
{\cal L}=\frac{G_F}{\sqrt{2}}\{
j_{V-A}^{\mu}J^{\dagger}_{V-A,\mu}+ \sum_{\alpha,\beta} 
 ^{'}\epsilon_{\alpha}^{\beta}j_{\beta} J^{\dagger}_{\alpha}\}
\ee
with the combinations of hadronic and leptonic Lorentz currents
$J_{\alpha}^{\dagger}=\bar{u} {\cal O}_\alpha d$ respectively 
$j_{\beta}= \bar{e} {\cal O}_\beta \nu$
of defined helicity. The operators 
${\cal O}_{\alpha,\beta}$ are defined as
\ba{10}
{\cal O}_{V-A}=\gamma^{\mu}(1-\gamma_5)\nn \\
{\cal O}_{V+A}=\gamma^{\mu}(1+\gamma_5)\nn \\
{\cal O}_{S-P}=(1-\gamma_5)\nn \\
{\cal O}_{S+P}=(1+\gamma_5)\nn \\
{\cal O}_{T_L}=\frac{i}{2}[\gamma_{\mu},\gamma_{\nu}](1-\gamma_5)\nn \\
{\cal O}_{T_R}=\frac{i}{2}[\gamma_{\mu},\gamma_{\nu}](1+\gamma_5).
\ea 
The prime indicates 
the sum runs over all contractions allowed by 
Lorentz--invariance,
except for $\alpha=\beta=V-A$. 
Note that all currents have been scaled relative to the strength 
of the ordinary ($V-A$) interaction. 

%
%CM_1: New addition on Lepton Number Violation
%
The effective Lagrangian given in eq. \rf{1} represents the most general 
low-energy 4-fermion charged-current interaction allowed by Lorentz 
invariance. The interpretation of the effective couplings 
$\epsilon^{\alpha}_{\beta}$, however, depend on the specific particle 
physics model. Nevertheless one realizes the following general 
feature. Using only the SM fermion fields\footnote{The use of 
only SM neutrinos does not really imply a loss of generality here, 
see the discussion on $V+A$ currents below.} 
and working in the 
Majorana basis for the neutrinos ($\nu := \nu_L + \nu_L^C$) it 
is easily seen that all currents involving operators proportional 
to ($1+\gamma_5$) violate 
lepton number by two units, i.e. the corresponding 
$\epsilon^{\alpha}_{\beta}$ must also be lepton-number violating. 
(Such LNV $\epsilon^{\alpha}_{\beta}$ are easily 
found, an example is given by R-parity violating supersymmetry 
treated in ref. \cite{Hir95,Hir96,Paes98}.)
%
%CM_1: END
%

The double beta decay amplitude is proportional to the time-ordered 
product of two effective Lagrangians (see Fig. 1): 

\ba{2}
T({\cal L}_{(1)} {\cal L}_{(2)})=\frac{G_F^2}{2}
T\{j_{V-A}J^{\dagger}_{V-A}j_{V-A}J^{\dagger}_{V-A}\nn \\
  + \epsilon_{\alpha}^{\beta}j_{\beta} J^{\dagger}_{\alpha}j_{V-A}
J^{\dagger}_{V-A} 
+ \epsilon_{\alpha}^{\beta}\epsilon_{\gamma}^{\delta} 
j_{\beta} J^{\dagger}_{\alpha} j_{\delta} 
J^{\dagger}_{\gamma}\}.
\ea
While the first term (contribution (a) in Fig. 1) corresponds to the 
standard model (SM) like neutrino exchange,
and
the 3rd term (contribution (c) in Fig. 1), which is quadratic in
$\epsilon$ can be neglected, only
the 2nd term (contribution (b))
is phenomenologically interesting. For this term one has to
consider two general cases:

\noindent
{\it 1)} The leptonic SM $V-A$ current meets a left--handed non SM current 
   $j_{\beta}$ with $\beta=S-P,T_L$. For this contribution the neutrino 
   propagator is
\be{12}
\sim P_L\frac{q^{\mu}\gamma_{\mu}+m_{\nu}}{q_{\mu}q^{\mu} - m_{\nu}^2}P_L=
\frac{m_{\nu}}{q_{\mu}q^{\mu} - m_{\nu}^2} 
\ee
   with the usual left-- and right--handed projectors 
   $P_{L/R}=\frac{1 \mp \gamma_5}{2}$. This expression
   is proportional to the unknown neutrino Majorana mass 
   $m_{\nu}\leq \sim 0.5 eV$, for which 
   no lower bound exists. Therefore no limits on the 
   corresponding  
   parameters $\epsilon_{\alpha}^{\beta}$ can be derived.

\noindent
{\it 2)} The leptonic SM $V-A$ current meets a right--handed non SM current
   $j_{\beta}$ with $\beta=S+P,V+A,T_R$. For this contribution the 
  neutrino propagator is
\be{13}
\sim P_L\frac{q^{\mu}\gamma_{\mu}+m_{\nu}}{q_{\mu}q^{\mu} - m_{\nu}^2}P_R=
\frac{q^{\mu}\gamma_{\mu}}{q_{\mu}q^{\mu} - m_{\nu}^2} 
\ee
which is proportional to the neutrino momentum 
$q_{\mu}\simeq p_F \simeq 100 MeV$ with the nuclear 
Fermi momentum $p_F$, and thus will produce stringent limits on
corresponding 
$\epsilon_{\alpha}^{\beta}$.

Taking these considerations into account, we are left with three interesting 
contributions discussed in the following 
section. 
With the present half-life limit of the Heidelberg--Moscow 
experiment $T_{1/2}^{0\nu\beta\beta}>1.2 \cdot 10^{25} y$ \cite{Kla97} 
and 
considering only one $\epsilon_{\alpha}^{\beta}$ at a time (evaluation
"on axis")
\be{t12}
[T_{1/2}^{0\nu\beta\beta}]^{-1}=(\epsilon_{\alpha}^{\beta})^2 G_{0k} |ME|^2
\ee
where $G_{0k}$ denotes the phase space factors given in \cite{Doi85} 
and $|ME|$ the nuclear 
matrix elements discussed in the following. Note that evaluating "on axis",
compared to the arbitrary evaluation, 
neglects interference terms of the different contributions. Although this is 
expected to be a small effect (see below), 
this remains to be discussed in the next step. 

\section{Calculational details and limits}
\subsection{SM meets $j_{V+A}J^{\dagger}_{V+A}$ and $j_{V+A}
J^{\dagger}_{V-A}$}
This contribution has been considered already in the context of left--right 
symmetric models \cite{Doi85,qrpa,Hir96c}. 
For sake of completeness we repeat the updated results of \cite{qrpa}
here in our notation:
$\epsilon^{V+A}_{V+A}< 7.9 \cdot 10^{-7}$, 
$\epsilon^{V+A}_{V-A}< 4.9 \cdot 10^{-9}$ for the full calculation with
arbitrary evaluation. Using s--wave approximation and "on axis" evaluation
as was done in this work the limits are reduced to
$\epsilon^{V+A}_{V+A}< 7.0 \cdot 10^{-7}$, 
$\epsilon^{V+A}_{V-A}< 4.4 \cdot 10^{-9}$.
This example confirms the expectation that these assumptions will only 
slightly (less than 10 \%) affect the result.

%
%CM_2: 
%
It is worthwhile discussing the following subtlety in interpretation 
when comparing our ansatz and the work of \cite{Doi85}. Doi et 
al. \cite{Doi85} calculate the decay amplitude, writing down the 
Lagrangian of an explicitly left-right symmetric model. Therefore, 
their calculations do not only treat SM neutrinos, but contain also 
right-handed neutrinos. Nevertheless, the derivation of the decay 
amplitude for $0\nu\beta\beta$ decay is the same in both calculations, 
only some 
care is required when going from one notation to the other. For example, 
in the lepton-number violating $\langle \lambda \rangle$, defined 
in \cite{Doi85} as

\be{deflambdadoi}
\langle \lambda \rangle = \lambda \sum'_{j}U_{ej}V_{ej}
\ee
where $\lambda \approx (m_{W_L}/m_{W_R})^2$, LNV is due to the 
product of mixing matrices $\sum'_{j}U_{ej}V_{ej}$, which vanishes 
identically if LNV goes to zero. Thus, our (LNV) $\epsilon^{V+A}_{V+A}$ 
corresponds to $\langle \lambda \rangle$ of Doi et al. \cite{Doi85}, 
(and {\it not} to $\lambda$). This example shows, that the 
use of only SM neutrinos in the derivation does not imply a loss 
of generality in our results, if the source of LNV of the particle 
physics model under consideration is carefully identified. 
%
%CM_2: END
%

\subsection{SM meets $j_{S+P}J^{\dagger}_{S+P}$ and $j_{S+P}
J^{\dagger}_{S-P}$}
Using s-wave approximation for the outgoing electrons and some assumptions 
according to \cite{tom91,Hir96} 
one gets 
\be{me1}
ME_{S+P}^{S+P}=-ME_{S-P}^{S+P}
=-\frac{F^{(3)}_{P}(0)}{R m_e G_A}\Big({\cal M}_{T^{'}}+\frac{1}{3}
{\cal M}_{GT^{'}}\Big).
\ee
The phase space factor is defined
\ba{phasspac}
G_{01}=\frac{(G_F G_A)^4 m_e^4}{{32} \pi^5 (m_e R)^2 ln2}
\int F_0(Z,\epsilon_1)p_1 \epsilon_1 F_0(Z,\epsilon_2)p_2\epsilon_2\nn \\ 
\delta(\epsilon_1 +\epsilon_2 +M_f -M_i)
d\epsilon_1 d\epsilon_2
\ea
and the matrix elements (summation over nucleons $a,b$ is suppressed) are
\ba{matr01}
{\cal M}_{GT^{'}}&=&
\langle 0_f^+| h_{R}(\vec{\sigma_a}
\vec{\sigma_b})\tau^+_a \tau^+_b| 0_i^+ \rangle \\
{\cal M}_{T^{'}}&=&\langle 0_f^+| h_{T^{'}}\{(\vec{\sigma_a}\hat{\vec{r}}_{ab})
(\vec{\sigma_b}\hat{\vec{r}}_{ab})
\nn \\
&&
-\frac{1}{3}(\vec{\sigma}_a
\vec{\sigma}_b)\} \tau^+_a \tau^+_b| 0_i^+ \rangle.
\ea
$h_R$ and $h_{T^{'}}$ are neutrino potentials defined as
\ba{nepot01}
h_R&=&\frac{2}{\pi}\frac{R^2}{m_P}\int_0^\infty dq q^4
\frac{j_0(qr_{ab})f^2(q^2)}{\omega(\omega+\overline{E})}, \\
h_{T^{'}}&=&\frac{2}{\pi}\frac{R^2}{m_P}\int_0^\infty dq q^2
\frac{f^2(q^2)}{\omega(\omega+\overline{E})}
\{q^2 j_0(qr_{ab})\nn \\
&&
-3 \frac{q}{r_{ab}} j_1(qr_{ab})\}. 
\ea
Here $R$ denotes the nuclear radius, $m_P$ the proton mass,
$\epsilon_i$ and $p_i$ are electron 
energies and momenta, $F_0(Z,\epsilon_i)$ and is the Fermi function.
Further $\omega=\sqrt{q^2+m_{\nu}^2}$, $q=|\vec{q}|$,
$\hat{r}=\vec{r}/r$ and $j_k(qr)$ are 
spherical Bessel functions. $(\omega+\overline{E})^{-1}$ is the energy 
denominator of the perturbation theory. The form factors 
$F^a_i(0)=F^a_i(q^2)/f(q^2)$ with $f(q^2)=(1+q^2/m_A^2)^{-2}$ 
($m_A^2=0.85$ GeV) 
have been calculated in the MIT bag model in \cite{adl}, $G_A \simeq 1.26$
and $G_V \simeq 1$.

Inserting the numerical value of the matrix elements
${\cal M}_{GT^{'}}$ and ${\cal M}_{T^{'}}$ (see \cite{Hir96} and 
Tab. 1), one derives 
$\epsilon^{S+P}_{S+P},\epsilon^{S+P}_{S-P}<1.1 \cdot 10^{-8}$.  

\subsection{SM meets $j_{T_{R}}J^{\dagger}_{T_{R}}$ and $j_{T_{R}}
J^{\dagger}_{T_{L}}$}
In the tensor part the decay rate depends on the phase space $G_{01}$ and
new matrix elements not considered
in the literature. 

For the hadronic $T_R$ contribution one gets under the assumptions 
used above 
\be{mat02}
ME^{T_R}_{T_R}=
- \alpha_1 \frac{2}{3} {\cal M}_{GT^{'}}
+\alpha_1 {\cal M}_{T^{'}}
\ee
with
\ba{T01nagneu}
\alpha_1&=&\frac{4 T_1^{(3)}(0)G_V(1-2m_P(G_W/G_V))}{G_A^2 R m_e}.
\ea
Again $T_1^{(3)}(0)$  has been taken from \cite{adl}
and the strength of the induced weak magnetism 
$(G_W/G_V)=\frac{\mu_P-\mu_n}{2m_P}\simeq \frac{-3.7}{2m_P}$ is obtained by the CVC
hypothesis. The involved nuclear matrix elements have been calculated in the
QRPA--approach of \cite{qrpa,qrpa2}.
Inserting the values obtained for the special case of $^{76}$Ge 
(see Tab. 1) yields $\epsilon^{T_{R}}_{T_{R}}< 1.7 \cdot 10^{-9}$.

For the hadronic $T_L$ contribution in leading order of $(1/m_P)$ one finds
\be{TL}
ME^{T_R}_{T_L}=
\alpha_2 {\cal M}_{F^{'}} 
-\alpha_3 \Big( {\cal M}_{T^{''}}+ \frac{1}{3}{\cal M}_{GT^{''}}
\Big)
\ee
with
\ba{MTss}
{\cal M}_{F^{'}}&=&\langle 0_f^+| h_R \tau_a^+ \tau_b^+ | 0_i^+ \rangle\\
{\cal M}_{GT^{''}}&=& 
\langle 0_f^+| h_{R\omega}(\vec{\sigma_a}
\vec{\sigma_b})\tau^+_a \tau^+_b| 0_i^+ \rangle \\
{\cal M}_{T^{''}}&=& 
\langle 0_f^+|\omega
h_{T^{''}}\{(\vec{\sigma_a}\hat{\vec{r}}_{ab})
(\vec{\sigma_b}\hat{\vec{r}}_{ab}) \nn \\
&&-\frac{1}{3}(\vec{\sigma}_a
\vec{\sigma}_b)\} \tau^+_a \tau^+_b| 0_i^+ \rangle.
\ea

The neutrino potentials are
\ba{MTsss}
h_R\omega&=&\frac{2}{\pi}\frac{R^3}{m_P}\int_0^\infty dq q^4
\frac{j_0(qr_{ab})f^2(q^2)}{(\omega+\overline{E})}, \\
h_{T^{''}}&=&\frac{2}{\pi}\frac{R^3}{m_P}\int_0^\infty dq q^2
\frac{f^2(q^2)}{(\omega+\overline{E})}
\{q^2 j_0(qr_{ab})\nn \\
&&
-3 \frac{q}{r_{ab}} j_1(qr_{ab})\}. 
\ea
and
\ba{T2}
\alpha_2&=&\frac{4 (2{\hat T}_2^{(3)}(0)-T_1^{(3)}(0))G_V}{G_A^2 R m_e},\\
\alpha_3&=&\frac{4 T_1^{(3)}(0)(G_P/G_A)}{G_A R^2 m_e}.
\ea
Here $\hat{T}_2^{(3)}(0)$ has been taken from \cite{adl}$,
G_P/G_A=2m_P/m_{\pi}^2$ and $\omega$ is the neutrino energy.
Again the matrix elements have been calculated in the model of 
\cite{qrpa,qrpa2}.
A limit of
$\epsilon^{T_{R}}_{T_{L}}< 7.3\cdot 10^{-10}$
has been obtained. 
Factors $R$
have been arbitrarily absorbed into the definition ${\cal M}_{GT^{''}}$ and 
${\cal M}_{T^{''}}$ to get dimensionless 
quantities. One should notice that the corresponding factor included in 
$\alpha_3$ compensates this choice.

\section{Conclusion}
We have presented a general parameterization for the long range part of the
neutrinoless double beta decay rate in terms of effective couplings. 
The resulting bounds are summarized in Tab. 2. 
Combined with the short range part and contributions of derivative 
couplings, this parameterization will give the double beta decay constraints
for arbitrary lepton number violating theories beyond the SM. 
The next 
step should include these contributions 
and discuss interference terms.

\section*{Acknowledgement}
M.H. would like to acknowledge support by the European Union's 
TMR program under grant ERBFMBICT983000.

\newpage

\begin{figure}
\epsfysize=45mm
\epsfbox{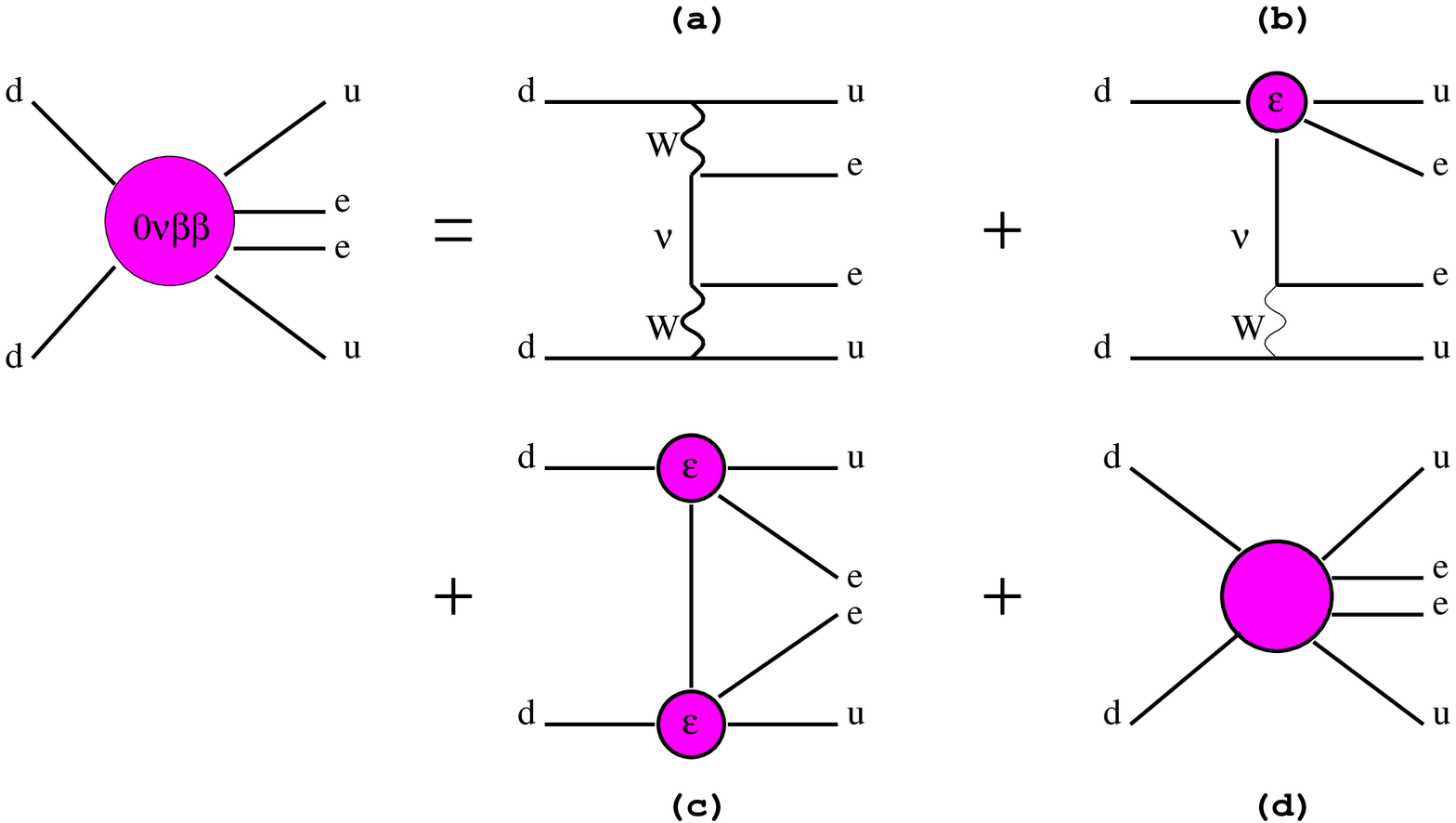}
\caption{\it Feynman graphs of the general double beta rate: 
The contributions a) - c)
correspond to the long range part, the contribution d) is the short range
part (to be discussed elsewhere).}
\label{1} 
\end{figure}
\vspace*{1cm}

 \begin{table}
 \begin{tabular}{l|l}
 \hline
 ${\cal M}_{GT^{'}}$&2.95 \\
 \hline
 ${\cal M}_{F^{'}}$&-0.663 \\
 \hline
 ${\cal M}_{GT^{''}}$&8.78 \\
 \hline
 ${\cal M}_{T^{'}}$&0.224 \\
 \hline
 ${\cal M}_{T^{''}}$&1.33 \\
 \hline
 \end{tabular}
 \vspace*{3mm}
  \caption{\it Relevant Nuclear Matrix Elements calculated in pn-QRPA}
 \end{table}

\begin{table}
 \begin{tabular}{l|l}
 \hline
 $\epsilon^{V+A}_{V-A}$ & $4.4 \cdot 10^{-9}$ \\
 \hline
 $\epsilon^{V+A}_{V+A}$ & $7.0 \cdot 10^{-7}$  \\
 \hline
 $\epsilon^{S+P}_{S-P}$ & $1.1 \cdot 10^{-8}$  \\
 \hline
 $\epsilon^{S+P}_{S+P}$ &  $1.1 \cdot 10^{-8}$ \\
 \hline
 $\epsilon^{TR}_{TL}$ &  $6.4 \cdot 10^{-10}$  \\
 \hline
 $\epsilon^{TR}_{TR}$ & $1.7 \cdot 10^{-9}$  \\
 \hline
 \end{tabular}
 \vspace*{3mm}
  \caption{\it Limits on effective $B-L$ violating couplings evaluated 
``on axis''}
 \end{table}


\begin{thebibliography}{99}


\bibitem{Kla97}
H.V. Klapdor--Kleingrothaus, in \cite{ring}


\bibitem{HM97}
L. Baudis {\it et al.} (Heidelberg--Moscow collab.),
Phys. Lett. {\bf B 407} (1997) 219

\bibitem{Hir95}
M. Hirsch, H.V. Klapdor--Kleingrothaus, S.G. Kovalenko, Phys. Rev. Lett. 
{\bf 75} (1995) 17, M. Hirsch, H.V. Klapdor--Kleingrothaus, S. Kovalenko, 
Phys. Rev. {\bf D 53} (1996) 1329

\bibitem{Hir96}
 M. Hirsch, H.V. Klapdor--Kleingrothaus, S.G. Kovalenko,
Phys. Lett. {\bf B 372} (1996) 181, Erratum: Phys. Lett. {\bf B 381} 
(1996) 488 

\bibitem{Paes98}
H. P\"as, M. Hirsch,  
H.V. Klapdor--Kleingrothaus, in preparation

\bibitem{Hir97}
M. Hirsch, H.V. Klapdor--Kleingrothaus, S.G. Kovalenko, 
Phys. Lett. {\bf B 398} (1997) 311; 
Phys. Rev. {\bf D 57} (1998) 1947;  
contribution in \cite{ring}

\bibitem{ring}
H.V. Klapdor--Kleingrothaus, H. P\"as (Eds.), Proc. Int. Conf.
{\it ``Beyond the Desert - Accelerator- and Non-Accelerator Approaches''},
Castle Ringberg, Germany, 1997


\bibitem{Hir96b}
M. Hirsch, H.V. Klapdor--Kleingrothaus, S.G. Kovalenko,
Phys. Lett. {\bf B 378} (1996) 17 and Phys. Rev. {\bf D 54}
(1996) R4207


\bibitem{Hir96c}
M. Hirsch, H.V. Klapdor--Kleingrothaus, O. Panella, 
Phys. Lett. {\bf B 374} (1996) 7


\bibitem{Pan97}
O. Panella, in \cite{ring}

\bibitem{Tak97}
E. Takasugi, in \cite{ring}

\bibitem{Doi85}
M. Doi, T. Kotani, E. Takasugi, Progr. Theor. Phys. Suppl. {\bf 83} (1985) 1

\bibitem{tom91}
T. Tomoda, Rep. Progr. Phys. {\bf 54} (1991) 53 

\bibitem{adl}
S. Adler {\it et al.}, Phys. Rev. {\bf D 11}, (1975) 3309

\bibitem{qrpa}
K. Muto, E. Bender, H.V. Klapdor, Z. Phys. {\bf A 334} (1989) 177,187;

\bibitem{qrpa2}
A. Staudt, K.Muto, H.V. Klapdor--Kleingrothaus, 
Europhys. Lett. {\bf 13} (1990) 31;
M. Hirsch, K.Muto, T.Oda, H.V. Klapdor--Kleingrothaus, Z. Phys. {\bf A 347} 
(1994) 151 

\end{thebibliography}
\end{document}